\newcommand{\caproman}[1]{\uppercase\expandafter{\romannumeral#1}}

\newcommand{\BGmail}{bgottsch\,@\,fas.harvard.edu}

\documentclass{article}

\usepackage{amsmath}
\usepackage{latexsym}
\usepackage{graphicx}
\usepackage{authblk}
\usepackage{hyperref}

\title{Comparison of Geant4 multiple Coulomb scattering models with theory for radiotherapy protons}

\author[1,3,4]{Anastasia Makarova }
\author[2]{Bernard Gottschalk}
\author[1]{Wolfgang Sauerwein}
\affil[1]{NCTeam, Department of Radiation Oncology, Medical Faculty, University of Duisburg-Essen, Essen 45122, Germany}
\affil[2]{Harvard University Laboratory for Particle Physics and Cosmology, 18 Hammond St., Cambridge, Massachusetts 02138, USA}
\affil[3]{Global Station for Quantum Medical Science and Engineering, Global Institution for Collaborative Research and Education (GI-CoRE), Hokkaido University, Sapporo 060-8648, Japan}
\affil[4]{Department of Radiation Oncology, Stanford University, Stanford, California 94305-5847, USA}

\begin{document}

\maketitle
\begin{center}
\href{mailto:anastasia.makarova@uni-due.de}{anastasia.makarova@uni-due.de}, \href{mailto:\BGmail}{\BGmail}
\end{center}

\begin{abstract}
\noindent
Usually, Monte Carlo models are validated against experimental data. However, models of multiple Coulomb scattering (MCS) in the Gaussian approximation are exceptional in that we have theories which are probably more accurate than the experiments which have, so far, been done to test them. In problems directly sensitive to the distribution of angles leaving the target, the relevant theory is the Moli\`ere/Fano/Hanson variant of Moli\`ere theory \cite{moliere2,bethe,mcsbg}. For transverse spreading of the beam in the target itself, the theory of Preston and Koehler \cite{preston,transport2012} holds.

Therefore, in this paper we compare Geant4 simulations, using the Urban and Wentzel models of MCS, with theory rather than experiment, revealing trends which would otherwise be obscured by experimental scatter. For medium-energy (radiotherapy) protons, and low-$Z$ (water-like) target materials, Wentzel appears to be better than Urban in simulating the distribution of outgoing angles. For beam spreading in the target itself, the two models are essentially equal.
\end{abstract}

\section{Introduction}

Monte Carlo (MC) simulations are the gold standard for dose calculations in proton radiotherapy. Geant4 (G4) is perhaps the most popular MC, particularly if we take into account packages based on it such as GATE \cite{Jan2004} and TOPAS \cite{Perl2012}. It is therefore important that the physics models in G4 be validated. The present paper focuses on multiple Coulomb scattering (MCS) in the Gaussian approximation. 

A 1993 paper \cite{mcsbg} (hereinafter Go93) contains measurements of the rms projected Gaussian angle $(\theta_x)_{rms}$ at 158.6\,MeV incident energy for 14 materials, with 115 material/thickness combinations in total. Go93 also summarizes the formulas of Moli\`ere theory, covering low-$Z$ targets, thick targets, compounds and mixtures, and the Gaussian approximation.

The CERN URL
\begin{center}
\texttt{http:\-//vnivanch.\-web.\-cern.ch/\-vnivanch/\-verification/verification/\-electromagnetic/\-MSCP/\-geant4-10-02-patch-01/}
\end{center}
links to a site comprising 14 graphs showing, for six G4 models, the difference in $(\theta_x)_{rms}$ between G4 simulations and the Go93 experimental data. A summary graph shows $\chi^2/N$ for each material and G4 setup (our Fig.\,\ref{fig:CERNchisq} is similar). The machinery to produce these graphs is described in a 2013 note by Schwarz \cite{Schwarz2013}; the most recent graphs were evidently generated around 15FEB2016. We discuss this work at greater length below. Suffice it to say here that the Wentzel model agrees best with experiment, as we will also find. However, because of experimental scatter, the CERN graphs give little insight into the dependence of either the Wentzel or Urban models on target material and/or thickness.

Fortunately, in the special case of MCS we have the luxury of an accurate theory free of adjustable parameters. Comparison with measurements in Go93 on many different target materials and thicknesses shows the Moli\'ere/Fano theory to be accurate to better than 1\% on the average. It appears to break down only when the target is thicker than = 97\% of the mean proton range. 
In this paper we take advantage of that by provisionally assuming that the Moli\`ere/Fano/Hanson variant appropriate to the Gaussian approximation (hereinafter `Hanson') is ground truth insofar as MCS is concerned. We justify that assumption by comparing the Go93 experimental data to Hanson. We then compare, also with Hanson, G4 computations of MCS using the Wentzel and Urban models. Further analysis reveals the dependence of the Wentzel and Urban models on target material and thickness, trends which would otherwise be obscured by experimental error.

Go93 is a `target/drift' experiment: the spread in projected angle $(\theta_x)_{rms}$, introduced by MCS in a target, is converted into a transverse spread $x_{rms}$ by a drift region, approximated by a large air gap, in which the additional scattering is small. Given the effective thickness of the air gap, $(\theta_x)_{rms}$ may be inferred from measured $x_{rms}$. 

Another class of experiments might be termed `beam spreading'. The transverse spreading $x_{rms}$ of an incident pencil beam {\em in the target itself} is measured as a function of depth. Such experiments are also regarded as tests of MCS models \cite{Grevillot2010,Matysiak2013} and also obey an accurate, experimentally tested theory with no adjustable parameters. We include them for completeness, though our primary emphasis is on $(\theta_x)_{rms}$.

A side issue that will arise is the dependence of computed $(\theta_x)_{rms}$ (for thick targets) on the range-energy relation of protons in the target material. That relation (judging by differences between standard tables) is uncertain to 1--2\%, affecting independently both the Hanson computation and the G4 simulations. We will show that the numerical effect on either is small.

\section{Methods}

\subsection{Experiment}

For completeness, we summarize the Go93 experiment. A well collimated 158.6\, MeV proton beam was directed onto the target and transverse scans were taken with a small Si diode 100\,cm distant from the upstream target face. Each scan was fit with a Gaussian on a constant background to find $x_{rms}$. Target-out `air' scans were taken and analyzed similarly and their $x_{rms}$ was subtracted in quadrature to correct for beam size, scattering in air, and detector size. Finally, corrected $x_{rms}$ was converted to $(\theta_x)_{rms}$ using an effective drift length that took into account the effective scattering point in the target. 

Go93 compared $(\theta_x)_{rms}$ with Highland's formula, a parameterization of Moli\`ere/Bethe/Hanson theory \cite{highland, highlandcorr}. We will, instead, use Moli\`ere/Fano/Hanson theory directly, which should be slightly better. In all, Go93 studied 14 target materials of potential interest in proton radiotherapy, spanning the periodic table. Target thicknesses ranged from very thin to somewhat greater than the mean proton range.

\subsection{Theory}\label{sec:theory}

\subsubsection{Target/Drift Experiments}

In the Gaussian approximation the 2D distribution $f(\theta)$ of polar angle $\theta$ is given by
\begin{equation}\label{eqn:f}
f(\theta)\,\theta\,d\theta\,d\phi\;=\;\frac{1}{2\pi\,\theta_0^2}\;e^{\textstyle{-\frac{1}{2}\left(\frac{\theta}{\theta_0}\right)^2}}\,\theta\,d\theta\,d\phi
\end{equation}
where
\begin{equation}
\theta_0=(\theta_x)_{rms}=\theta_{rms}/\sqrt{2}
\end{equation}
 Eq.\,\ref{eqn:f} is valid to $\theta\approx2.5\,\theta_0$, where the Moli\`ere single scattering tail becomes appreciable \cite{transport2012}. This Gaussian region contains about 96\% of the protons and therefore dominates proton radiotherapy dose calculations.  

As Hanson et al. \cite{hanson} first observed, the best Gaussian fit to Moli\`ere theory is obtained, not by merely using the first (Gaussian) term in Moli\`ere's expansion of $f(\theta)$, but by letting 
\begin{equation}
\theta_0=\chi_c\sqrt{B-1.2}/\sqrt{2}
\end{equation}
where $\chi_c$ is Moli\`ere's characteristic single scattering angle and $B$ is his reduced target thickness. For the Moli\`ere/Fano/Hanson computation of $\theta_0$ we find these quantities using Moli\`ere's rather than Bethe's form of the theory ($Z^2$ rather than $Z(Z+1)$) and using the Fano correction for low-$Z$ targets (see Go93). The appropriate formulas are embodied in Fortran program LOOKUP \cite{BGware}. (In a minor improvement, LOOKUP uses cubic spline interpolation, rather than a polynomial fit, to interpolate range-energy tables.) We used the default MIXED range-energy table, namely ICRU\,49 \cite{icru49} except for Nylon, Zn and brass which are Janni\,82 \cite{janni82}.

For thin targets (negligible energy loss) $\theta_0$ depends only on the initial value of
\begin{equation}\label{eqn:pv}
pv\;=\;\frac{(T/mc^2)+2}{(T/mc^2)+1}\;T
\end{equation}
where $p$, $v$, $T$ and $mc^2$ are proton momentum, speed, kinetic energy and rest energy. (In the clinical regime $3\le T\le300$\,MeV the fraction multiplying $T$ ranges from 2 to 1.76 so $pv$ is roughly twice the kinetic energy.) 

For thick targets, the integrals in Moli\`ere theory depend on the relation of $pv$ to depth in the target, and the range-energy relation comes into play. To estimate this effect we replaced MIXED, the range-energy table one would use nowadays, by Janni\,66 \cite{janni66}, the tables (now outdated) used by Go93. The largest change in $\theta_0$, for near-stopping Pb, was 3.1\%, and for most material/thickness combinations it was far smaller. We will perform an analogous test in the G4 simulations. Table\,\ref{tab:results} lists typical values of $\theta_0$\,(Hanson) for reference.

\subsubsection{Beam Spreading Experiments}
Unlike Moli\`ere theory, which is complicated, beam spreading in a homogeneous slab follows just two rules first derived and tested experimentally by Preston and Koehler \cite{preston}. They hold for protons or heavier ions stopping in any material at any incident energy. A modern derivation is given in \cite{transport2012}.

The first rule is that the rms transverse spread $\sigma_x(R)$ at end-of-range $R$ is proportional to range. The constant of proportionality is
\begin{equation}\label{eqn:sigxoR1}
\frac{\sigma_x(R)}{R}\;=\;\frac{E_s\,z}{2\,(pv)_\mathrm{R/2}}\sqrt{\frac{R}{X_S}}
\end{equation}
which despite appearances is very nearly independent of $R$. $E_s=15.0$\,MeV, $z$ is the particle charge number, $pv$ is evaluated at the $T$ value corresponding to $R/2$, and $X_S$ is the scattering length \cite{scatPower2010} of the material. In Lexan, for instance, $\sigma_x(R)=0.0210\,R$. Values of $\sigma_x(R)/R$ and $\rho X_S$ for many other materials are given in \cite{transport2012}.

The second rule is that, at any lesser depth $z<R$,
\begin{equation}\label{eqn:PKuniversal}
\frac{\sigma_x(z)}{\sigma_x(R)}\;=\;\left[2\,(1-t)^2\ln\left(\frac{1}{1-t}\right)
  +\,3\,t^2-2\,t\right]^{1/2}\hbox{\quad,\quad}t\;\equiv\; z/R
\end{equation}
which, with Eq.\,\ref{eqn:sigxoR1}, completely describes beam spreading in a homogeneous slab. Eqs.\,\ref{eqn:sigxoR1} and \ref{eqn:PKuniversal} assume an ideal incident beam, so measurements of $\sigma_x(z)$ must be corrected for initial beam size, divergence and emittance. These can be significant in beams designed for pencil beam scanning, but will not concern us. We will simply assume an ideal beam.

\subsection{Geant4 Setup}

\subsubsection{Physics}\label{sec:physics}

We used G4 10.02, the latest release at the time of writing. Electromagnetic physics was based on the \texttt{G4\-Em\-Standard\-Physics\-\_option4} physics constructor class providing the most accurate models available in standard and low energy categories. The physics constructor was modified to allow different models and different parameters of the MCS process to be activated for protons. Stopping power tables were limited to the range $0-200$\,MeV and the number of bins was increased to 50 per decade according to the recommendations of Grevillot et al. \cite{Grevillot2010}. Other parameters, such as the step limitation function for the stopping process and other physical processes, were left at their default values.

Two MSC models, \texttt{G4\-Urban\-Msc\-Model} and \texttt{G4\-Wentzel\-VI\-Model} were tested with the default step limitation and lateral displacement parameters. Then, the impact of those parameters on the results was investigated. \texttt{G4\-Urban\-Msc\-Model} provides three step limitation options: \texttt{f\-Mi\-ni\-mal}, \texttt{f\-Use\-Safety} and \texttt{fUse\-Distance\-To\-Boundary}, while \texttt{G4\-Wentzel\-VI\-Model} has only the default step limitation and a \texttt{fUse\-Distance\-To\-Boundary} option. Special attention should be paid to the way the step limitation, lateral displacement options and other parameters of MCS and other electromagnetic processes are defined. Static class \texttt{G4\-Em\-Parameters} was added recently to the G4 library specifically for this purpose. Its methods \texttt{Set\-Mu\-Had\-Lateral\-Dis\-placement()} and \texttt{Set\-Msc\-Mu\-Had\-Step\-Limit\-Type()} control switching the lateral displacement and type of step limitation algorithm for the hadronic MCS process on or off. By default, these parameters are set to \texttt{false} and \texttt{f\-Minimal} respectively.

The \texttt{G4\-Coulomb\-Scattering} process was not tested. Though it provides accuracy comparable with solving the diffusion equation, it is far too slow and thus inapplicable in proton therapy calculations.

\subsubsection{Material Properties}

Table \ref{tab:materials} lists properties of the fourteen materials in this study. Names beginning with G4 indicate G4 default compositions and $I_\mathrm{G4}$ (mean excitation energy) values. For the others, we used densities and compositions from Go93 and computed $I_\mathrm{G4}$ by the internal G4 procedure. The mean projected range $R_\mathrm{G4}$ corresponds to the peak of differential fluence found in an auxiliary simulation. 

As noted earlier, for thick targets $\theta_0$ depends on the range-energy relation. Generally, when reconciling an MC with (say) an experimental Bragg peak, one can fine-tune either the incident energy or $I$. Here, however, we {\em must} use $I$ since $\theta_0$ depends directly on the incident energy via Eq.\,\ref{eqn:pv} even for thin targets. To change $I$ by a reasonable amount we can adjust it to reproduce, via G4, some well-known range-energy table other than the one that follows from G4 defaults. Somewhat arbitrarily, we choose Janni 1966 \cite{janni66}, the first comprehensive tables used in proton radiotherapy and the ones used in Go93.

Accordingly, for each simulation with $I_\mathrm{G4}$, we performed a second with $I_\mathrm{adj}$ adjusted to yield a range $R_\mathrm{adj}$ closely matching the range $R_\mathrm{Janni}$ from \cite{janni66} as given in Go93. These quantities are also given in Table\,\ref{tab:materials} as is the percent difference between $R_\mathrm{adj}$ and $R_\mathrm{G4}$. That reflects the difference between two plausible range-energy relations for an arbitrary assortment of fourteen materials. 

\subsubsection{Scoring and Analysis}\label{sec:scoring}

Unlike \cite{Schwarz2013} we did {\em not} (even approximately) simulate the Go93 experiment. Instead, a point mono-energetic mono-directional 158.6\,MeV proton source was placed in front of the material slab and 1\,M protons were traced from the source to the last step in the slab. For maximum efficiency, the polar angle $\theta$ of particles emerging from the last step, weighted by $1/\theta$, was scored in an annulus of radius $\theta$ and bin width $d\theta$ using the \texttt{G4\-Csv\-Analysis\-Manager} class.
That histogram was then fitted with a Gaussian (cf. Eq.\,\ref{eqn:f}) to find $\theta_0$.

\subsection{Graphs and Trend Analysis}
Let the {\em percent deviation} of e.g. experiment from Hanson theory be defined as
\begin{equation}
D_\mathrm{EH}\;\equiv\;100\times\left(\;\frac{\theta_0(E)}{\theta_0(H)}-1\right)
\end{equation}
where, if $D>0$, the quantity under test (E) is greater than ground truth (H). We plot $D_\mathrm{EH}$, $D_\mathrm{UH}$ and $D_\mathrm{WH}$ to exactly the same scales to facilitate comparison.  The abscissa (target mass thickness in g/cm$^2$) is logarithmic because of the known behavior of MCS with target thickness (see Go93).

At fixed energy, only the dependence of $\theta_0$ on target material and target thickness remain to be explored. To summarize the compliance of experiment, Urban and Wentzel to Hanson theory, we fit the data in Figs. \ref{fig:MCS_Exp}$-$\ref{fig:MCS_Wentzel} with straight lines. We exclude near stopping targets ($>0.9\times$\,mass range) where Moli\`ere theory fails because range straggling destroys the relation between $pv$ and depth. MCS in this region is of minor importance in proton radiotherapy, the residual range being so small that the proton direction hardly matters.

\begin{figure}[p] 
  \centering
  \includegraphics[width=0.8\textwidth]{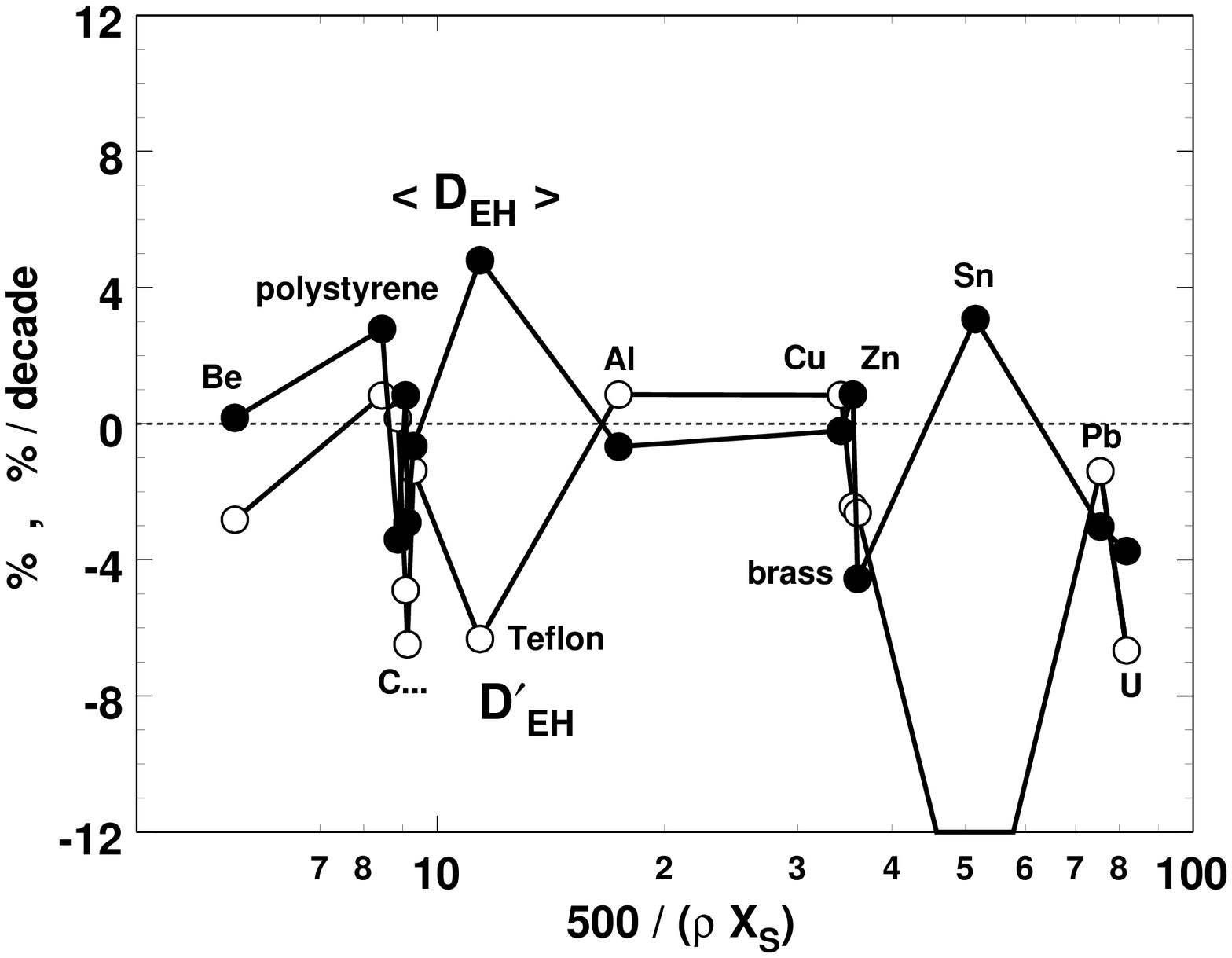}
  \caption{Fitted experimental results. The horizontal scale, (500\,g/cm$^2$)/(mass scattering length), is arbitrary. Filled circles are mean $D_\mathrm{EH}$ (\%); open circles are slope $D'_\mathrm{EH}$ (\%/decade). `C$\ldots$' stands for C, Lexan, Nylon and Lucite in that order.}
  \label{fig:DEH}
\end{figure}

\begin{figure}[p] 
  \centering
  \includegraphics[width=0.8\textwidth]{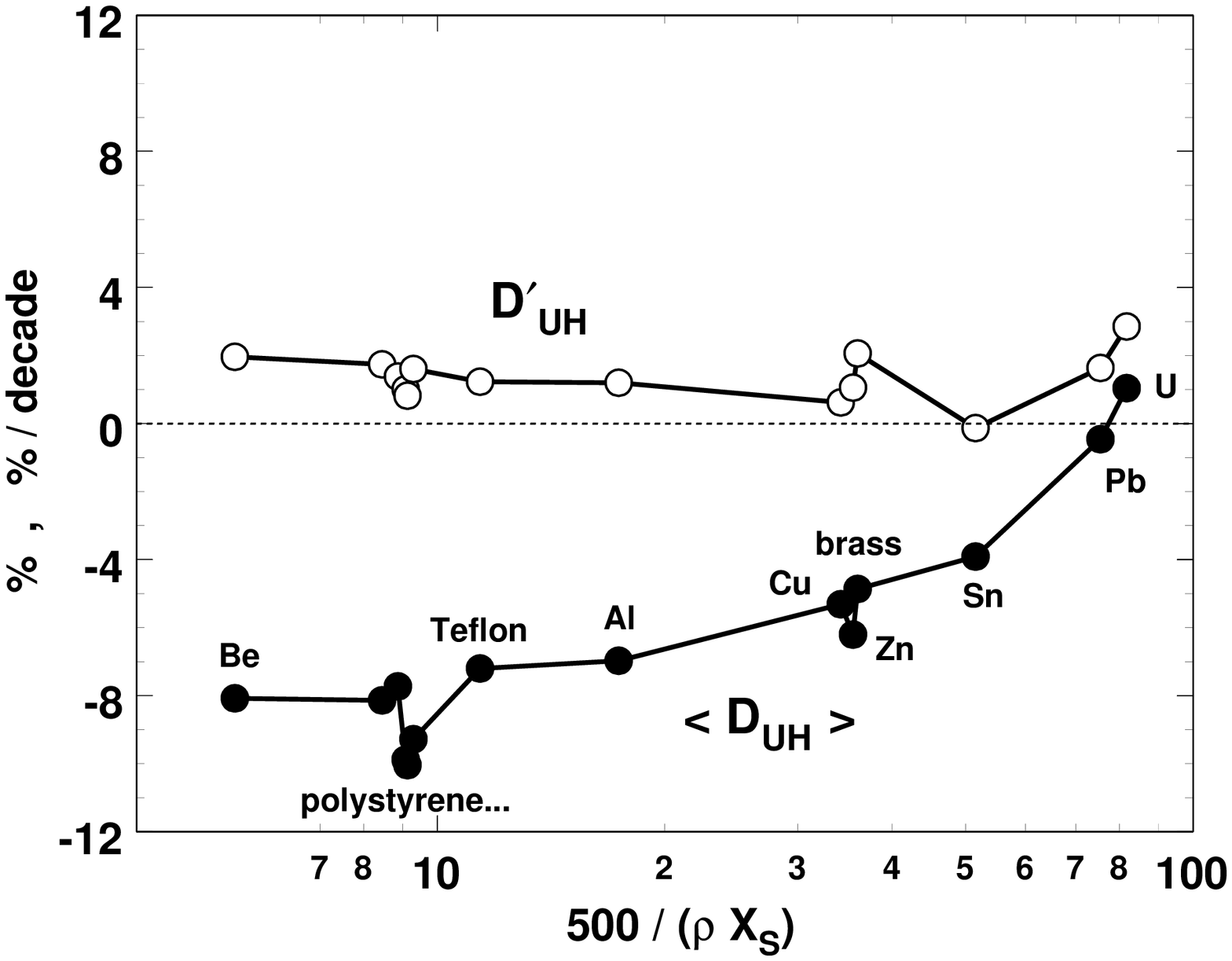}
  \caption{Fitted Urban results.  The horizontal scale, (500\,g/cm$^2$)/(mass scattering length), is arbitrary. Filled circles are mean $D_\mathrm{UH}$ (\%); open circles are slope $D'_\mathrm{UH}$ (\%/decade).  `polystyrene$\ldots$' stands for polystyrene, C, Lexan, Nylon and Lucite in that order.}
  \label{fig:DUH}
\end{figure}

\begin{figure}[p] 
  \centering
  \includegraphics[width=0.8\textwidth]{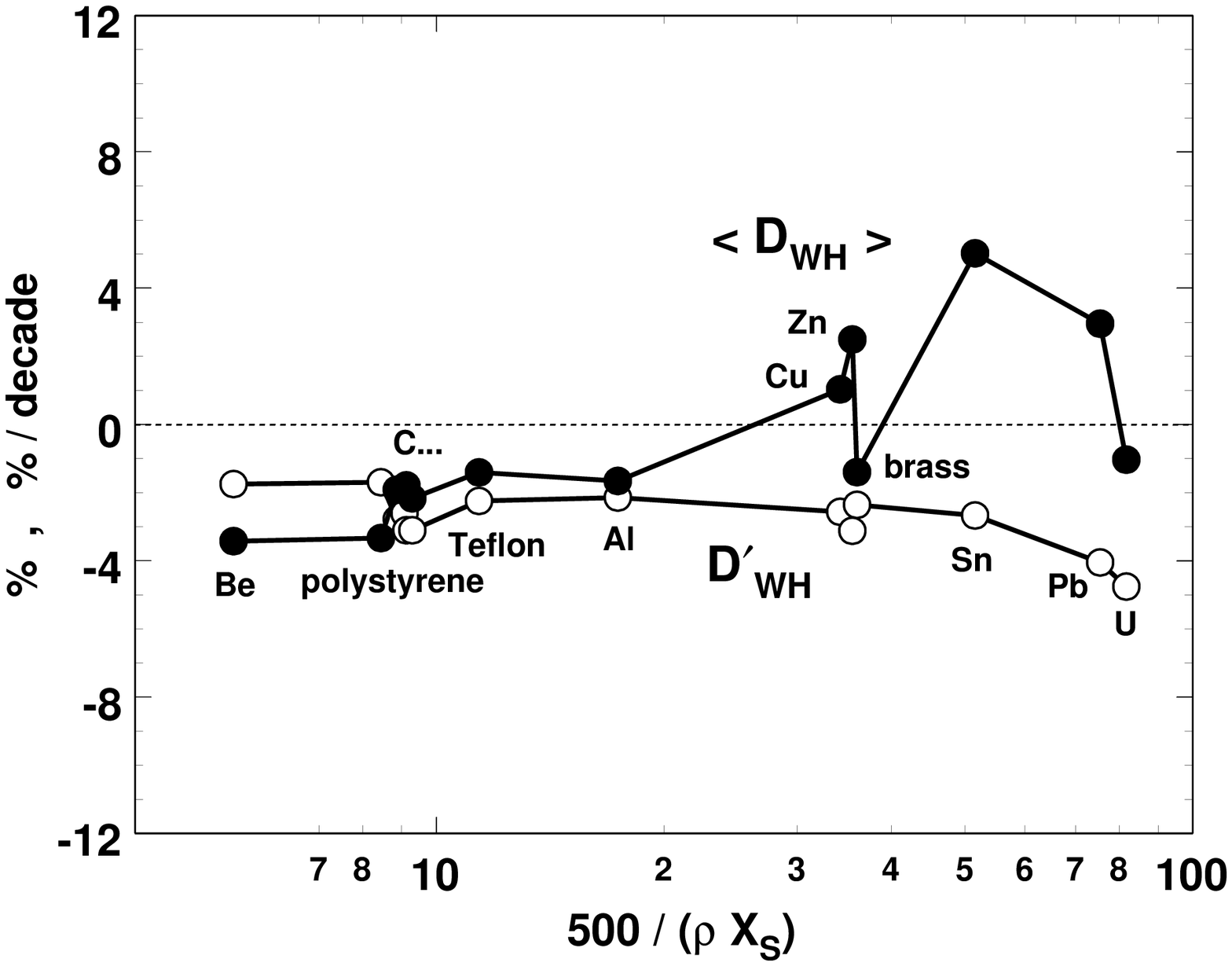}
  \caption{Fitted Wentzel results.  The horizontal scale, (500\,g/cm$^2$)/(mass scattering length), is arbitrary. Filled circles are mean $D_\mathrm{WH}$ (\%); open circles are slope $D'_\mathrm{WH}$ (\%/decade). `C$\ldots$' stands for C, Lexan, Nylon and Lucite in that order.}
  \label{fig:DWH}
\end{figure}

Finally we plot the slope $D'$ (thickness dependence, \%/decade) and mean value $<\!\!D\!\!>$ (material dependence, \%) of the fitted lines for E, U and W (Figs.\,\ref{fig:DEH}$-$\ref{fig:DWH}). Again, we use exactly the same scales to facilitate comparison.

\section{Results}\label{sec:results}

Table\,\ref{tab:results} gives, for selected points, the target material, g/cm$^2$, and measured $\theta_0$ from Go93, the computed $\theta_0$\,(Hanson) from LOOKUP using the MIXED range-energy table, and finally $\theta_0$\,(Urban) and $\theta_0$\,(Wentzel) from the G4 simulations using the G4 default $I$ values. 

\begin{figure}[t] 
  \centering
  \includegraphics[bb=0 0 1330 1669,width=5.67in,height=7.11in,keepaspectratio]{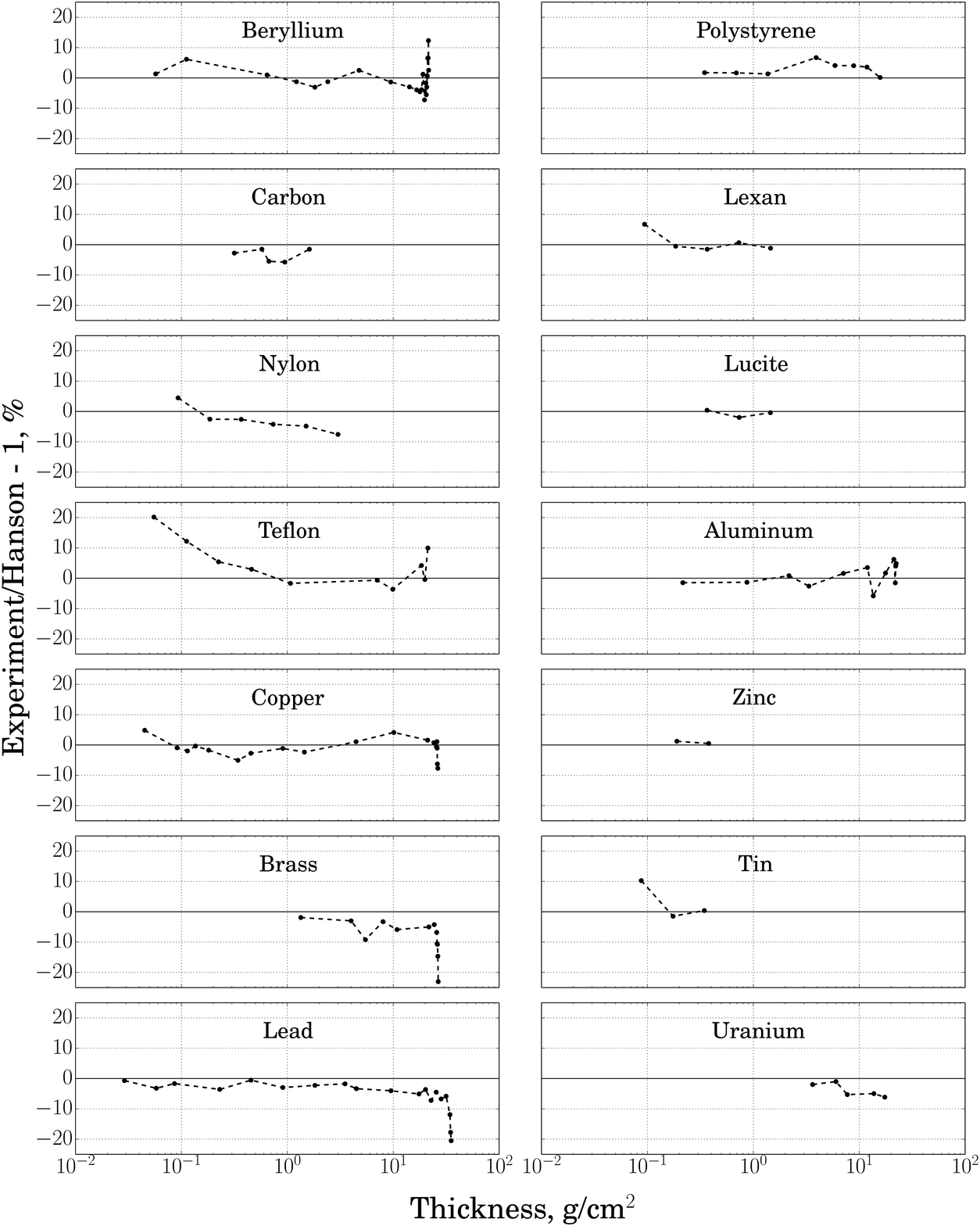}
  \caption{Comparison of experimental data from Go93 with `Hanson', the Moli\`ere/Fano/Hanson theory computed using LOOKUP and the MIXED range-energy table.}
  \label{fig:MCS_Exp}
\end{figure}

\begin{figure}[t] 
  \centering
  \includegraphics[bb=0 0 1329 1669,width=5.67in,height=7.12in,keepaspectratio]{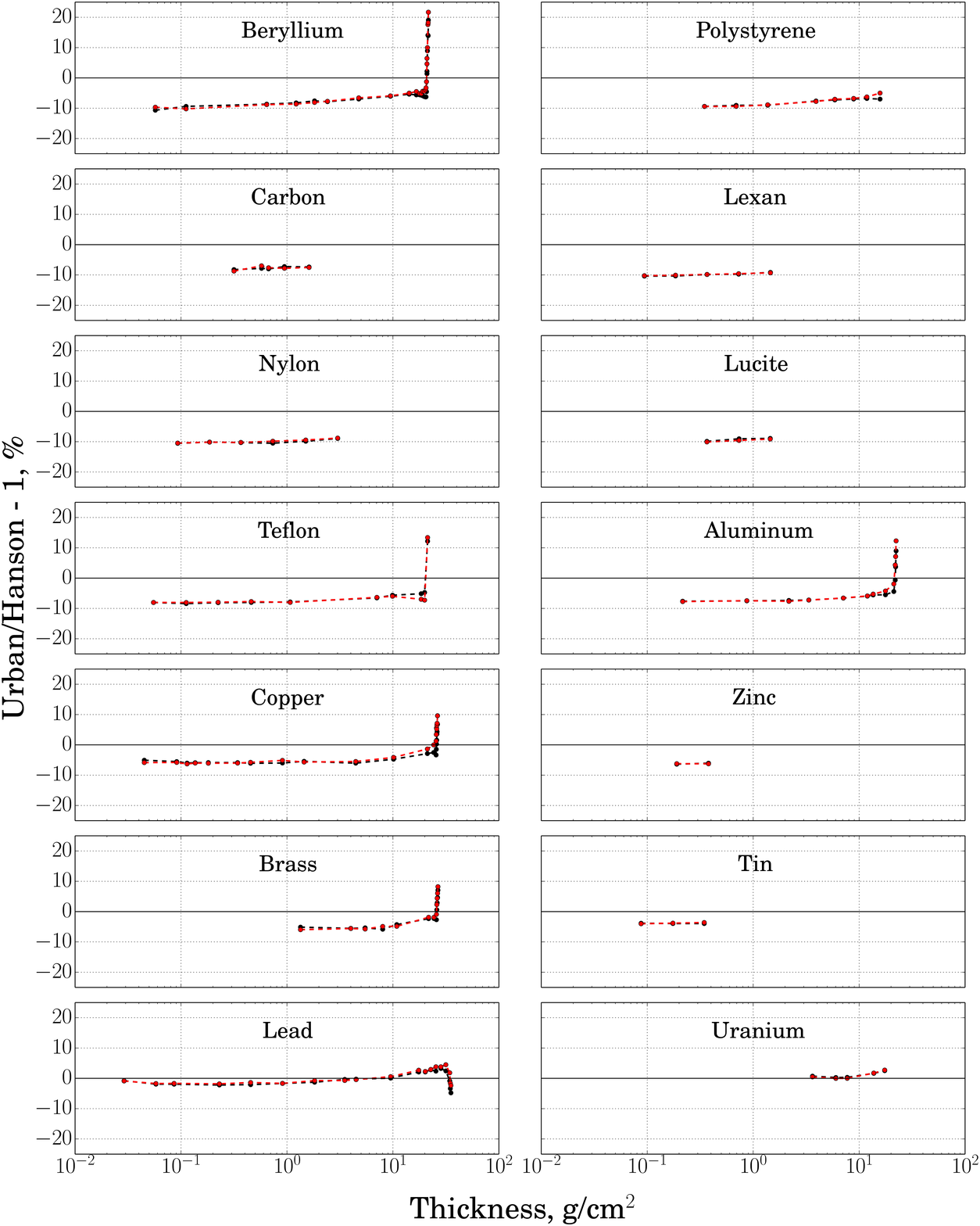}
  \caption{Comparison of the G4/Urban simulation with Hanson. Black points: simulation with G4 default values of $I$. Red points: simulation with $I$ tuned to fit the Janni66 range-energy tables \cite{janni66}.}
  \label{fig:MCS_Urban}
\end{figure}

\begin{figure}[t] 
  \centering
  \includegraphics[bb=0 0 1329 1669,width=5.67in,height=7.12in,keepaspectratio]{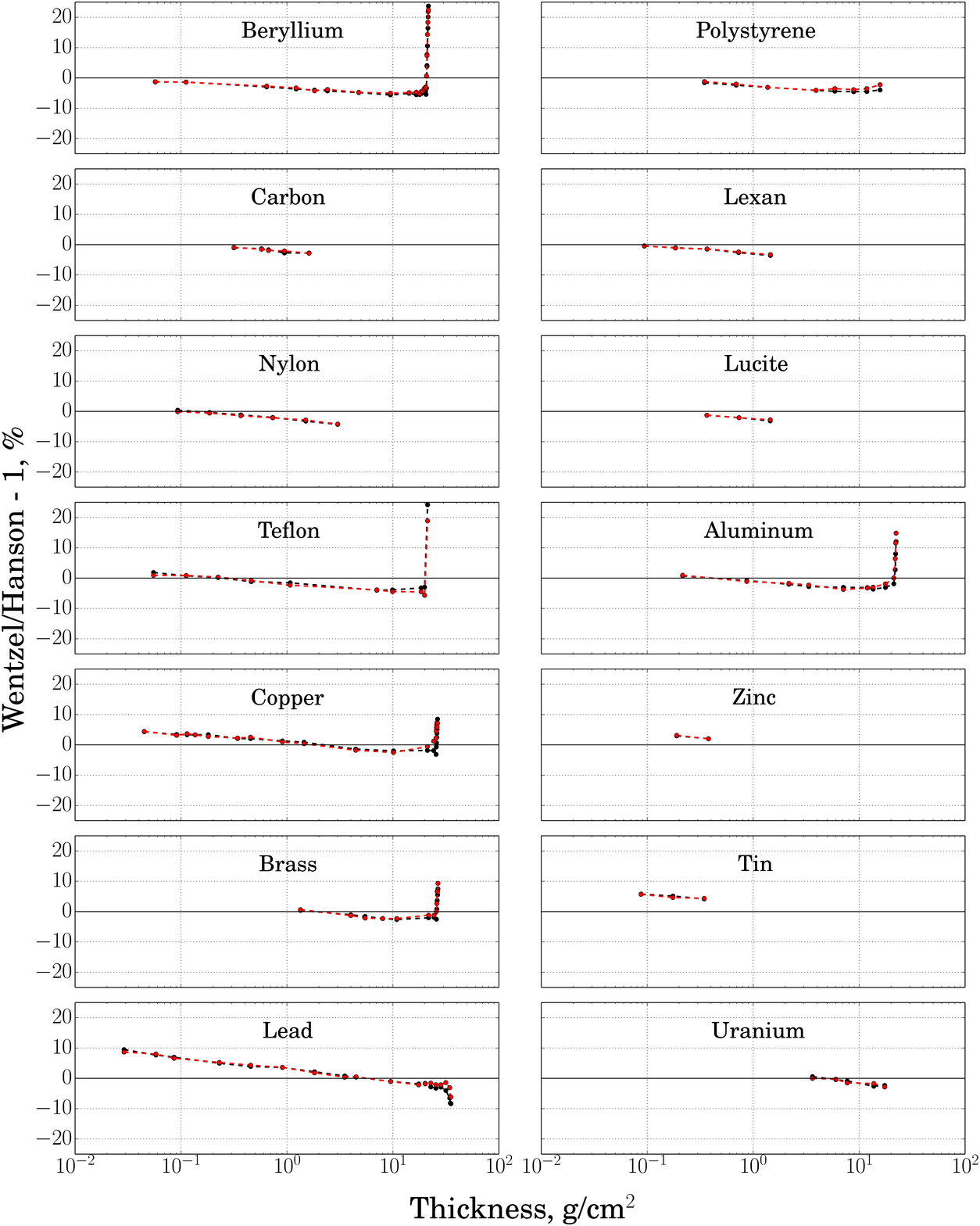}
  \caption{Comparison of the G4/Wentzel simulation with Hanson. Black points: simulation with G4 default values of $I$. Red points: simulation with $I$ tuned to fit the Janni66 range-energy tables \cite{janni66}.}
  \label{fig:MCS_Wentzel}
\end{figure}

Percent deviations of $\theta_0$\,(exptl), $\theta_0$\,(Urban) and $\theta_0$\,(Wentzel) from $\theta_0$\,(Hanson) are shown in Figs.\,\ref{fig:MCS_Exp}, \ref{fig:MCS_Urban} and \ref{fig:MCS_Wentzel} respectively. $1\sigma$ experimental errors (not shown) were taken from Go93.

Fig.\,\ref{fig:MCS_Exp}, taking into account the experimental error, shows that Hanson theory indeed describes the measurements with the possible exception of the thickest Pb and U points where theory may be some 4\% high (or experiment 4\% low) as already noted in Go93. 

Figs.\,\ref{fig:MCS_Urban} and \ref{fig:MCS_Wentzel} show the comparison of of Geant4 simulation with Urban and Wentzel models with Hanson theory. Typical behavior for both models is a deviation from theory which is nearly linear in the logarithm of target thickness, has a small positive or negative slope, and some average offset from 0. MC statistical errors are small and are already implied by the non-smoothness of the lines with dots bigger than error bars.

Fig.\,\ref{fig:DEH} quantifies the trends seen in Fig.\,\ref{fig:MCS_Exp}. Teflon and Sn are obvious outliers, almost certainly due to experimental error given the much better agreement of neighboring materials. In particular Be, Al and Cu, for each of which a full range of thicknesses was measured, agree with theory very well on average. 


Figs.\,\ref{fig:DUH} and \ref{fig:DWH} quantify and summarize the trends seen in Figs.\,\ref{fig:MCS_Urban} and \ref{fig:MCS_Wentzel}. Averaged over target thickness, the Urban model is $\approx8\%$ low for low-$Z$ targets, smoothly approaching $\approx0\%$ for Pb and U. The variation with $\log_{10}$(target thickness) is roughly linear, with a slope of $1\%-2\%$/decade.

By contrast, the Wentzel model is $\approx4\%$ low for low-$Z$ targets, agrees with theory at midrange, and is $\approx4\%$ high for high-$Z$ targets. Material dependence is noticeably less smooth than the Urban model. Thickness dependence is slightly greater than the Urban model and opposite in sign, say $-2\%$ to $-4\%$/decade.

\section{Discussion}\label{sec:discussion}

Four other studies known to us test the G4 MCS model. Only one \cite{Grevillot2010} is published.

\subsection{CERN Web Site}

This site, already mentioned, explores six G4 configurations. The only documentation appears to be the note by Schwarz \cite{Schwarz2013} based on which there are two major differences with the present work. 

First and foremost, G4 is compared with experimental measurements rather than theory. 

Second, \cite{Schwarz2013} describes a partial simulation of the Go93 experiment, unlike our method which merely scored $\theta$ of protons emerging from the target. In \cite{Schwarz2013} an ideal beam enters the target and proton hits are scored on a finely divided measuring plsne (to avoid detector size effects) 100\,cm downstream of the target entrance face. The intervening gap is void (to avoid scattering in air). A Gaussian is fitted to what is effectively the transverse fluence (rather than dose). To convert its $x_{rms}$ to $\theta_0$ the effective scattering point is calculated according to Go93.

This procedure seems somewhat roundabout compared to simply scoring angles emerging from the target, but it seems to account for everything except incident beam size. Beam size may explain why G4 is consistently low for the thinnest Be targets on this Web site.

\begin{figure}[p] 
  \centering
  \includegraphics[width=0.8\textwidth]{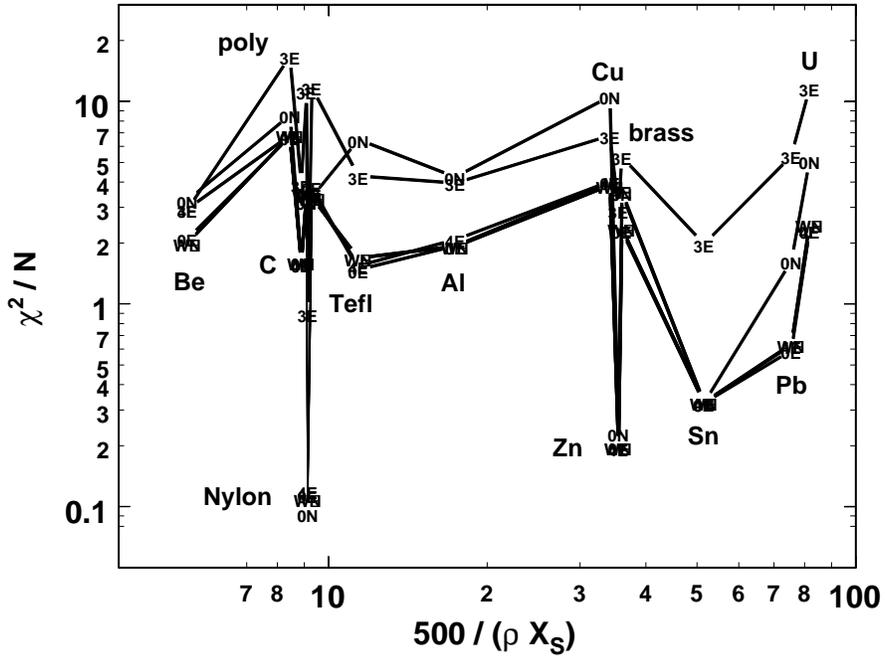}
  \caption{Summary of $\chi^2/N$ for CERN web site runs dated 15FEB2016. The abscissa is arbitrary; some materials are labeled for reference. All G4 configurations are prefaced `emstandard' and 3E = opt3 + elastic, 0N = opt0 + none, 4E = opt4 + elastic, 0E = opt0 + elastic, WE = WVI + elastic, WN = WVInoDisp + elastic. The last four are indistinguishable on this graph.}
  \label{fig:CERNchisq}
\end{figure}

Fig.\,\ref{fig:CERNchisq}, somewhat similar to a figure on the CERN site, summarizes the goodness-of-fit $\chi^2/N$ for the six configurations. Those labeled 3E and 0N are significantly worse than the four others, which are indistinguishable. Of those, 4E corresponds to our `Urban' and WE corresponds most closely to our `Wentzel'. The main point of Fig.\,\ref{fig:CERNchisq} is that $\chi^2/N$ alone is not a good way of evaluating the different models. It is too sensitive to the way individual points and errors happen to fall out.

\subsection{Fuchs et al.}

This poster presentation \cite{Fuchs2015} tests numerous G4 releases. Simulated $\theta_0$ is obtained directly, by scoring angles emerging from the target, or indirectly by back projecting dose profiles. The two methods agree. Those values of $\theta_0$ are then compared with the Go93 measurement for every material/thickness combination. The `10.1 mod EM Wentzel VI' release is found to be best, with an average error of only $-1.2\pm3.3\%$, in substantial agreement with our Fig.\,\ref{fig:DWH}. The range, $-17.9\%$ to $11.2\%$, is of course much larger owing to comparison with experiment rather than theory.

\subsection{Matysiak et al.}

In this poster presentation \cite{Matysiak2013} Matysiak et al. develop an MC tool to assess the accuracy of the Eclipse pencil beam model. 

First, to select the best G4 model, they consider spreading of an ideal beam in a 7.5\,cm water equivalent Lexan range shifter (RS) at six incident energies $120-226.7$\,MeV, each at four step sizes. (The RS thickness is 6.507\,cm.) Urban (option\,3) and Wentzel (option\,4) simulations are compared with $\sigma_x$ values from `analytical calculations using Moli\`ere scattering'. We computed our own $\sigma_x$ values, finding $\sigma_x(R)/R=0.0213$ from Eq.\,\ref{eqn:sigxoR1} and $\sigma_x(6.507\mathrm{\,cm})$ values from Eq.\,\ref{eqn:PKuniversal} consistently $8\%$ higher than Matysiak's. Therefore, over the energy range, Wentzel beam spreading in Lexan is either found to be $\approx4$ to $0\%$ high (Matysiak theory) or $\approx4$ to $8\%$ low (our theory). The Urban model is found to be step-size dependent and therefore not used further by Matysiak.

Having compared simulated beam spreading in a homogeneous slab with theory, Matysiak et al. proceed to compare a target/drift experiment with measurement, using the RS as the target. First, they determine the incident beam size, divergence and emittance by fitting measurements of the open beam in air. Next, inserting the RS and using those open beam parameters, they simulate and measure transverse fluence distributions at five locations along the beam axis covering a range of 35\,cm, fitting with Gaussians to find simulated and measured $\sigma_x$. G4/Wentzel is high by $\approx7\%$ at 120\,MeV improving to $\approx0\%$ at 226.7\,MeV. It is better in the $y$ direction than in the $x$ direction (presumably the bend plane).

Unlike beam spreading, the target/drift experiment comes close to a direct test of the G4 MCS model. Assuming the energy dependence to be largely due to sensitivity to beam parameters, and allowing for some experimental error, the Matysiak study is not inconsistent with our finding that Wentzel is $\approx2\%$ low in the neighborhood of Lexan cf. Fig.\,\ref{fig:DWH}.
 
\subsection{Grevillot et al.}

Grevillot et al. \cite{Grevillot2010} optimized GEANT4 settings for proton pencil beam scanning simulations using GATE. In the section relevant here, they found (their Figs.\,10 and 11) that GATE-simulated beam spreading in PMMA (Lucite) underestimated experiment by an energy dependent amount reaching 20\% at 210.56\,MeV incident. They measured $\sigma_x$ using EBT radiochromic film in a PMMA phantom. Quantitive dosimetry with radiochromic film is an exacting technique subject to nonlinear dose response, LET dependence and sensitivity to scanning technique \cite{AAPM63}. Indeed, they describe their own results as `preliminary' and `qualitative'.

\begin{figure}[p] 
  \centering
  \includegraphics[width=0.8\textwidth]{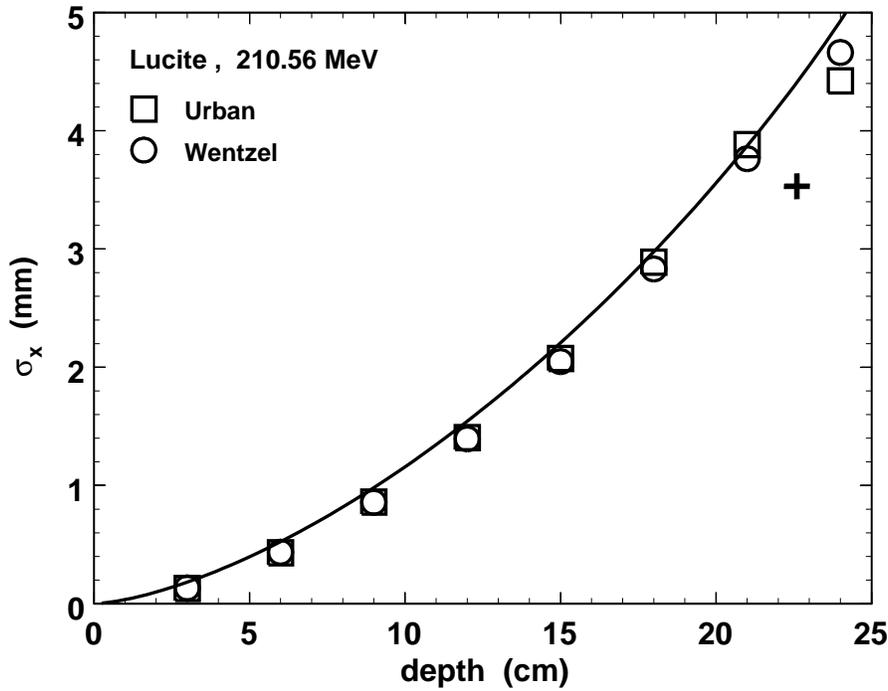}
  \caption{Beam spreading in Lucite (PMMA) at 210.56\,MeV incident. Line: theory of Preston and Koehler \cite{preston}. Points: simulation with G4 Urban and Wentzel models as described in text. For both, $<$error$>\,=-0.14$\,mm and $<$error$>$/$<\sigma_x>\,=-6.4$\,\%.
Cross: point at 22.6\,cm depth per Grevillot et al. \cite{Grevillot2010}}
  \label{fig:Grev2010}
\end{figure}

Even so, this paper invites the question whether G4 simulations of beam spreading in PMMA (and presumably, other water-like materials) could possibly be low by as much as 20\%. In Fig.\,\ref{fig:Grev2010} we compare a GMC simulation, with the settings described above, with the theory of Preston and Koehler \cite{preston} as summarized by Eqs.\,\ref{eqn:sigxoR1} and \ref{eqn:PKuniversal}. There is little difference between the Urban and Wentzel models. Both are poor near end-of-range and do well elsewhere. For both, average difference between MC and the theory  is $<$error$>\,=-0.14$\,mm and $<$error$>$/$<\sigma_x>\,=-6.4$\,\% including the last point. The possibility of a 20\% shortfall in G4 at 22.6\,cm (cross) is ruled out.

\section{Summary}

Measurements of the distribution of outgoing angles from a target test the MCS model of a Monte Carlo program and normally, models are validated directly against such experimental data. We have argued that, in the special case of MCS, theory may be taken as ground truth because it is free of adjustable parameters and agrees, on average, with a very large body of data. That reveals trends in the models that would otherwise be obscured by experimental scatter.

We have concentrated on the target/drift configuration, which measures outgoing angles. First, we justified the ground truth assumption by comparing experiment with the Moli\`ere/Fano/Hanson theory. We then compared G4 simulations, using the Urban and Wentzel models, with the same theory. 

Our Figs.\,\ref{fig:DUH} and \ref{fig:DWH} give Wentzel a slight advantage in proton radiotherapy where the materials of greatest interest are water-like. For the highest-$Z$ materials Urban is at least as good.

For completeness, we discussed beam-spreading experiments, also considered to be tests of the MCS model. Here the relevant theory is that of Preston and Koehler \cite{preston} as re-derived in \cite{transport2012} and summarized by our Eqs.\,\ref{eqn:sigxoR1} and \ref{eqn:PKuniversal}. G4 simulation of beam spreading in PMMA agrees with theory to a fraction of a millimeter or $6.4\%$ for both models, contradicting the finding of Grevillot et al. \cite{Grevillot2010}.

\section{Acknowledgments}

BG is indebted to the Physics Department of Harvard University for continuing support.

\clearpage
\bibliographystyle{unsrt}
\bibliography{biblio}

\begin{table}[p]
\centering
\caption{Selected results: experiment, theory and G4 simulations. Material, thickness and $\theta_0$\,(exptl) are from Go93. $\theta_0$\,(Hanson) is from Moli\`ere/Fano/Hanson theory (Sec.\,\ref{sec:theory}). $\theta_0$\,(Urban) and $\theta_0$\,(Wentzel) are from Gaussian fits to G4 runs (Secs.\,\ref{sec:physics}--\ref{sec:scoring}). First and last fitted points and some intermediate ones are given.\label{tab:results}}
\begin{tabular}[]{lccccc}
\noalign{\vspace{10pt}}
\hphantom{$\theta_0$ (Wentzel)}&\hphantom{$\theta_0$ (Wentzel)}&\hphantom{$\theta_0$ (Wentzel)}&\hphantom{$\theta_0$ (Wentzel)}&\hphantom{$\theta_0$ (Wentzel)}&\hphantom{$\theta_0$ (Wentzel)}\\
\multicolumn{1}{c}{Material}&Thickness&$\theta_0$\,(exptl)&$\theta_0$\,(Hanson)&$\theta_0$\,(Urban)&$\theta_0$\,(Wentzel)\\
&g/cm$^2$&mrad&mrad&mrad&mrad\\
\noalign{\vspace{5pt}}
Beryllium&  0.0572 & 0.993 & 0.980 & 0.876 & 0.967\\
&  1.820 & 6.394 & 6.596 & 6.096 & 6.331\\
&  20.313 & 43.848 & 44.601 & 41.798 & 42.589\\
\noalign{\vspace{8pt}}
Polystyrene&  0.347 & 3.346 & 3.289 & 2.980 & 3.237\\
&  15.751 & 42.031 & 41.973 & 39.039 & 40.304\\
\noalign{\vspace{8pt}}
Carbon&  0.316 & 3.084 & 3.172 & 2.911 & 3.139\\
&  1.616 & 7.728 & 7.846 & 7.268 & 7.630\\
\noalign{\vspace{8pt}}
Lexan & 0.094 & 1.762 & 1.651 & 1.480 & 1.643\\
 & 1.455 & 7.436 & 7.523 & 6.834 & 7.254\\
\noalign{\vspace{8pt}}
  Nylon & 0.093 & 1.727 & 1.653 & 1.479 & 1.659\\
&  3.010 & 10.656 & 11.529 & 10.499 & 11.035\\
\noalign{\vspace{8pt}}
  Lucite & 0.366 & 3.558 & 3.544 & 3.194 & 3.498\\
&  1.449 & 7.579 & 7.610 & 6.931 & 7.370\\
\noalign{\vspace{8pt}}
  Teflon & 0.055 & 1.626 & 1.353 & 1.244 & 1.378\\
&  1.072 & 6.918 & 7.037 & 6.484 & 6.928\\
&  19.908 & 64.003 & 64.274 & 61.234 & 62.358\\
\noalign{\vspace{8pt}}
  Aluminum & 0.216 & 3.534 & 3.587 & 3.314 & 3.613\\
&  2.173 & 13.104 & 12.995 & 12.038 & 12.733\\
&  21.245 & 87.103 & 81.996 & 78.369 & 80.453\\
\noalign{\vspace{8pt}}
  Copper & 0.045 & 2.204 & 2.102 & 1.995 & 2.193\\
&  1.450 & 14.327 & 14.671 & 13.875 & 14.799\\
&  24.250 & 118.561 & 117.658 & 114.698 & 115.435\\
\noalign{\vspace{8pt}}
  Zinc & 0.190 & 4.884 & 4.825 & 4.518 & 4.968\\
 & 0.379 & 7.131 & 7.096 & 6.667 & 7.240\\
\noalign{\vspace{8pt}}
  Brass & 1.342 & 14.120 & 14.394 & 13.655 & 14.451\\
&  24.398 & 115.851 & 120.982 & 118.213 & 118.697\\
\noalign{\vspace{8pt}}
  Tin & 0.0875 & 4.113 & 3.730 & 3.586 & 3.945\\
 & 0.345 & 8.106 & 8.074 & 7.756 & 8.411\\
\noalign{\vspace{8pt}}
  Lead & 0.029 & 2.304 & 2.320 & 2.302 & 2.540\\
&  0.907 & 16.093 & 16.585 & 16.309 & 17.179\\
&  31.566 & 175.421 & 186.292 & 190.846 & 178.819\\
\noalign{\vspace{8pt}}
  Uranium & 3.630 & 36.942 & 37.688 & 37.961 & 37.905\\
&  17.430 & 95.288 & 101.524 & 104.097 & 99.147
\end{tabular}
\end{table}

\begin{table}[p]
\centering
\caption{Material properties used in G4 runs. $I_\mathrm{G4},R_\mathrm{G4}$: default G4 mean ionization potential and mass range; $I_\mathrm{adj}$, $R_\mathrm{adj}$: the same with $I$ adjusted to match $R_\mathrm{Janni}$, the mass range stated in Go93 from 
a polynomial fit to \cite{janni66}; final column: deviation of $R_\mathrm{adj}$ from $R_\mathrm{G4}$.\label{tab:materials}}
\vspace{10pt}
\begin{tabular}[]{ lcccccccc }
  Material & \multicolumn{2}{c}{G4 material or}&$I_{G4}$&$R_{G4}$&$I_\mathrm{adj}$&$R_\mathrm{adj}$&$R_\mathrm{Janni}$&$R_\mathrm{adj}/R_\mathrm{G4}-1$\\
  & \multicolumn{2}{c}{g/cm$^3$, frac. wt.}& eV & g/cm$^2$ & eV & g/cm$^2$& g/cm$^2$&\%\\
\noalign{\vspace{8pt}}
   Beryllium & \multicolumn{2}{c}{G4\_Be} & 63.7 & 21.333 & 60.4 & 21.099 & 21.108&-1.10\\
   Polystyrene &\multicolumn{2}{c}{G4\_POLYSTYRENE} & 68.7 & 17.682 & 62.3 & 17.494 & 17.504&-1.06\\
Carbon & \multicolumn{2}{c}{G4\_C} & 81.0 & 19.513 & 74.3 & 19.278 & 19.270&-1.20 \\
\noalign{\vspace{5pt}} 
 Lexan & 1.20  & \begin{tabular}{c c} C & 0.741 \\ O & 0.185 \\ H & 0.074 \end{tabular} & 68.4 & 17.790 & 65.5 & 17.666 & 17.667&-0.70\\
\noalign{\vspace{5pt}} 
 Nylon & 1.13  & \begin{tabular}{c c} C & 0.549 \\ O & 0.244 \\ N & 0.107 \\ H & 0.100 \end{tabular}&64.8&17.250&62.1&17.190& 17.195&-0.35\\
\noalign{\vspace{5pt}} 
 Lucite & 1.20  & \begin{tabular}{c c} C & 0.600 \\ O & 0.320 \\ H & 0.081 \end{tabular} & 68.5 & 17.614 & 67.1 & 17.594 & 17.584&-0.11 \\
\noalign{\vspace{5pt}} 
 Teflon & \multicolumn{2}{c}{G4\_TEFLON} & 99.1 & 20.883 & 106.5 & 21.003 & 21.008&0.57\\
 Aluminum & \multicolumn{2}{c}{G4\_Al} & 166.0 & 22.401 & 155.6 & 22.155 & 22.158&-1.10\\
 Copper & \multicolumn{2}{c}{G4\_Cu} & 322.0 & 26.265 & 289.9 & 25.947 & 25.923&-1.21\\
 Zinc & \multicolumn{2}{c}{G4\_Zn} & 330.0 & 26.235 & 312.8 & 26.007 & 25.985&-0.87\\
\noalign{\vspace{5pt}} 
 Brass & 8.489  & \begin{tabular}{c c} Cu & 0.615 \\ Zn & 0.352 \\ Pb & 0.033 \end{tabular} & 333.7 & 26.439 & 319.8 & 26.325 & 26.345&-0.43\\
\noalign{\vspace{5pt}} 
 Tin & \multicolumn{2}{c}{G4\_Sn} & 488.0 & 30.678 & 437.2 & 30.188 & 30.159&-1.60\\
 Lead & \multicolumn{2}{c}{G4\_Pb} & 823.0 & 35.844 & 754.4 & 35.196 & 35.209&-1.81\\
 Uranium & \multicolumn{2}{c}{G4\_U} & 890.0 & 37.148 & 834.4 & 36.788 & 36.776&-0.97\\
\end{tabular}
\end{table}

\clearpage

\end{document}